\documentclass[10pt]{extarticle}
\usepackage{graphicx}
\usepackage{amssymb}
\def\beq{\begin{eqnarray}}
\def\eeq{\end{eqnarray}}
\usepackage[titletoc,toc,title]{appendix}

\usepackage{amsfonts}

\begin{document}

\title{Exact sum rules for quantum billiards of arbitrary shape }
\author{Paolo Amore \\
\small Facultad de Ciencias, CUICBAS, Universidad de Colima,\\
\small Bernal D\'{i}az del Castillo 340, Colima, Colima, Mexico \\
\small paolo.amore@gmail.com }

\maketitle

\begin{abstract}
We have derived explicit expressions for the sum rules of order one of the eigenvalues of the negative Laplacian
on two dimensional domains of arbitrary shape. Taking into account the leading asymptotic behavior of these 
eigenvalues, as given from Weyl's law, we show that it is possible to define sum rules that are finite,
using different prescriptions.
We provide the explicit expressions and test them on a number of non trivial examples, comparing the exact
results with precise numerical results.
\end{abstract}


\section{Introduction}

We consider the Helmholtz equation on a two dimensional region $\Omega$
\begin{eqnarray}
- \Delta \Psi_n = E_n \Psi_n \ \ \ ; \ \ \ n=1,2,\dots
\label{Helmholtz1}
\end{eqnarray}
where $E_n$ are the eigenvalues and $\Psi_n$ the eigenfunctions, obeying appropriate boundary conditions
on $\partial\Omega$~\footnote{In our discussion the boundary conditions we either be Dirichlet 
($\left. \Psi_n\right|_{\partial\Omega}=0$) or Neumannn  
($\left. \hat{n} \cdot \nabla\Psi_n\right|_{\partial\Omega}=0$, where $\hat{n}$ is the unit vector normal
to the border at each point).}.

Eq.~(\ref{Helmholtz1}) can be solved exactly only in special cases such as for the circle or the rectangle and
therefore one must rely on approximations for more general cases. For instance, if the domain $\Omega$ is a 
perturbation of a circle  (or of any other domain where the exact solutions are known), perturbation theory (PT) 
allows one to obtain explicit expressions for the eigenvalues and eigenfunctions of eq.~(\ref{Helmholtz1}) as a 
power series in the perturbation parameter; when the domain $\Omega$ is not a perturbation, the equation can still 
be solved numerically, for a limited portion of the spectrum, using the different techniques. 
Variational estimates can also be used to provide rigorous bounds on specific eigenvalues (in particular on the 
lowest eigenvalue).

The asymptotic behavior of the spectrum, on the other hand, is described by Weyl's law, which relates the counting function
$N(E) = \#\left\{E_n \leq E\right\}$ to the geometric properties of the domain
\begin{eqnarray}
N(E) = \frac{A}{4 \pi} E \mp \frac{L}{4\pi} \sqrt{E} + o(\sqrt{E}) \ \ \ ; \ \ \ E \rightarrow \infty
\end{eqnarray}
where $A$ and $L$ are respectively the area and perimeter of the domain. The $\mp$ sign refers to Dirichlet/Neumann 
boundary conditions.

A useful strategy in dealing with  Eq.~(\ref{Helmholtz1}) is to apply a conformal transformation, mapping the original region
into a suitable region, where a complete set of eigenfunctions of the Laplacian is known (the existence of this map is 
granted by Riemann's mapping theorem, although finding the explicit expression may be difficult). 

In this case the Helmholtz equation is trasformed to the equation (see for example ref.~\cite{Kuttler84})
\begin{eqnarray}
- \Delta \Psi_n = E_n \Sigma(x,y) \Psi_n  
\label{Helmholtz2}
\end{eqnarray}
where $\Sigma$ is a "density" related to the conformal transformation which maps $\Omega$ to the circle (or to any suitable 
region). The solutions to this equation can be approximated using either analytical or numerical methods: for example, 
when the map is a perturbation of the identity one can apply perturbation theory (see ref.~\cite{Amore10}), whereas in more
general cases, spectral (Rayleigh-Ritz) ~\cite{Robnik84} or pseudospectral (collocation) ~\cite{Amore10} methods can be used to obtain precise results.

Itzykson and collaborators ~\cite{Itzykson86} have first noticed that it is possible to obtain explicit (exact) expressions
for the sum rules of the eigenvalues of Eq.~(\ref{Helmholtz2}) 
\begin{eqnarray}
\zeta(s) = \sum_{p=1}^\infty \frac{1}{E_p^s} \ \ ; \ \ s=2,3, \dots
\label{intro_1}
\end{eqnarray}
calculating the traces of the appropriate operators, without the need of explicitly knowing the eigenvalues $E_n$.
Eq.(6) of ref.~\cite{Itzykson86} deals with the special case $s \rightarrow 1^+$, for which $\zeta(s)$ diverges, 
identifying the leading contributions. 
Kvitsinsky ~\cite{Kvitsinsky96} has later applied similar ideas to discuss the spectral sum rules of nearly circular domains,
discussing in particular the case of regular n-sided polygonal domains (note that the analysis of Refs.~\cite{Itzykson86} 
and ~\cite{Kvitsinsky96} is limited to Dirichlet boundary conditions).

Dittmar \cite{Dittmar02} has obtained explicit formulas for the sum rules of order two, both for Dirichlet and Neumann boundary 
conditions, for simply connected domains of the plane, using a conformal transformation of the original domain to the unit disk.
Examples of sums rules of order two for the cardioid and related domains are given in ref.~\cite{Dittmar11}.

More recently we have derived the general expressions for the spectral sum rules of inhomogenous strings and membranes (in one and two dimensions), 
of which Eq.~(\ref{Helmholtz2}) is a special case, for different boundary conditions (see refs.~\cite{Amore13A,Amore13B,Amore14}). 
In this way we were able to derive explicit expressions for the spectral sum rules of a circular sector and of a symmetric annulus with Dirichlet
boundary conditions.  The case of boundary conditions allowing a zero mode (which would correspond, for instance, to the circular annulus with Neumann 
boundary conditions),  specifically discussed in  ref.~\cite{Amore14}, is particularly delicate because of extra contributions that have to 
be correctly taken into account.

We refer the reader interested in other examples of spectral sum rules for related problems to the references cited in 
ref.~\cite{Amore13A,Amore13B, Amore14}.

The purpose of this paper is to extend our previous results to the calculation of sum rules of order one for two dimensional domains: 
we will show that, by following the adequate prescriptions, it is possible to define sum rules of order one which are finite, even 
though the expression (\ref{intro_1}) diverges at $s = 1$. We will define generalizations of eq.~(\ref{intro_1}) which are 
finite at $s=1$ and verify the analytical results with precise numerical estimates for a number of non-trivial examples.

The paper is organized as follows: in section \ref{sec:sum_rules} we discuss the procedures needed to obtain finite sum rules and 
obtain the corresponding explicit expressions in terms of a trace; in section \ref{sec:appl} we consider several examples, obtaining
the exact expressions for the sum rules of order one and comparing them with precise numerical estimates; in section \ref{sec:concl}
we draw our conclusions.

\section{Spectral sum rules}
\label{sec:sum_rules}

As we have discussed in our previous papers \cite{Amore13A,Amore13B, Amore14} it is possible to obtain an explicit formula
for the spectral sum rules 
\beq
Z(p) = \sum_{n} \frac{1}{E_n^p}  \ \ \ ; \ \ \ p=2,3,\dots
\label{spectral_sum}
\eeq
where $E_n$ are the eigenvalues of the Helmholtz equation on a finite two dimensional region 
(in the case of a string, i.e. in one dimensions, $p=1$ is also allowed), expressing 
$Z(p)$  as a trace in terms of a $p$-points "free" Green's function and of $p$ "densities", 
each evaluated at a different internal point (see  Eq.(65) of ref.~\cite{Amore13A}). 

Remarkably, the calculation of $Z(p)$, when expressed in this form, does not require the exact 
(and actually not even approximate) knowledge  of any of the eigenvalues $E_n$. 
The reader will find several examples of calculations of sum rules for one and two dimensional problems, with different boundary conditions,
in refs.~\cite{Amore13A,Amore13B, Amore14}. 

Let us now discuss the case of spectral sum rules of order one. It is straightforward to see that $Z(1)$ diverges 
for the two dimensional problem since Weyl's law implies 
\beq
E_n \approx \frac{4\pi n}{A} + \dots \ \ \ , \ \ \ n \rightarrow \infty 
\label{2_Weyl}
\eeq
where $A$ is the area of the domain where eq.(\ref{Helmholtz1}) is being solved. The divergence of eq.~(\ref{spectral_sum}) 
for $p=1$, manifests itself in the singular behavior of the Green's function, when the trace is taken.

To illustrate this point we will consider the disk, with either Dirichlet or Neumann boundary conditions at the border.

The Dirichlet Green's functions for the disk is reported  in ref.~\cite{Duffy} and it reads
\begin{eqnarray}
G^{(D)}(r,\theta,r',\theta') &=& - \frac{1}{4\pi} \log \left( r^2 + {r'}^2 -2 r r' \cos (\theta-\theta')\right) \nonumber \\ 
&+& \frac{1}{4\pi} \log \left( r^2 {r'}^2 + 1 -2 r r' \cos (\theta-\theta')\right) 
\end{eqnarray}
whereas the Neumann Green's function for the disk can be found in ref.~\cite{Kolokolnikov06} and it reads~\footnote{Note that this Green's function is the analogous of the "regularized Green's function" discussed in ref.~\cite{Amore14} for the rectangle.}
\begin{eqnarray}
G^{(N)}(r,\theta,r',\theta') &=& -\frac{\log \left(r^2-2 r r' \cos\left(\theta -\theta '\right)+\left(r'\right)^2\right)}{4 \pi } \nonumber \\
&-& \frac{\log \left(-2 r r' \cos \left(\theta -\theta '\right)+r^2 \left(r'\right)^2+1\right)}{2 \pi} \nonumber \\
&+& \frac{r^2}{4 \pi }+\frac{\left(r'\right)^2}{4 \pi } -\frac{3}{8 \pi }
\end{eqnarray}

To start with, we consider the Dirichlet Green's function for the disk, evaluated at two points infinitesimally close,
taking $r' = r + \eta \delta r$ and  $\theta' = \theta+ \eta \delta\theta$ and $\eta \rightarrow 0$:
\beq
G^{(D)}(r,\theta,r + \eta \delta r,\theta+ \eta \delta\theta)  \approx  \frac{\log \left(\left(1-r^2\right)^2\right)
-\log \left(\eta^2 \left({\delta r}^2+r^2 \ \delta \theta ^2 \right)\right)}{4 \pi } + O(\eta)
\label{2_GD}
\eeq 

The corresponding expression for $Z_D(1)$ on an arbitrary domain will formally read
\beq
Z_D(1) =  \lim_{\eta\rightarrow 0} \int_0^1 dr \int_0^1 d\theta  \ r \ G^{(D)}(r,\theta,r + \eta \delta r,\theta+ \eta \delta\theta)  \  \Sigma(r,\theta)
\label{2_ZD}
\eeq
where $\Sigma(r,\theta)$ is the "conformal" density of eq.~(\ref{Helmholtz2}). $Z_D(1)$ will diverge for $\eta \rightarrow 0$ because of the behavior in
eq.~(\ref{2_GD}).

A similar result holds for Neumann boundary conditions; in this case the Green's function behaves
\beq
G^{(N)}(r,\theta,r + \eta \delta r,\theta+ \eta \delta\theta)  \approx \frac{r^2}{2 \pi } -\frac{3}{8 \pi } 
-\frac{\log \left(\eta^2 \left({\delta r}^2+r^2 \delta \theta^2 \right)\right)+\log \left(\left(1-r^2\right)^2\right)}{4 \pi }
\label{2_GN}
\eeq

As proved in ref.~\cite{Amore14} for the case of boundary conditions allowing a zero mode the corresponding expression for $Z_N(1)$ on an arbitrary 
domain in two dimensions is
\beq
Z_N(1) &=&  \lim_{\eta\rightarrow 0} \int_0^1 dr \int_0^1 d\theta  \ r \ G^{(N)}(r,\theta,r + \eta \delta r,\theta+ \eta \delta\theta)    \Sigma(r,\theta) \nonumber \\
&+& \frac{\int_0^1 dr \int_0^{2\pi}d\theta \int_0^1 dr' \int_0^{2\pi}d\theta'  \ r r' \ G^{(N)}(r,\theta,r',\theta') 
\Sigma(r,\theta)\Sigma(r',\theta')}{\int_0^1 dr \int_0^{2\pi}d\theta \ r \ \Sigma(r,\theta)}
\label{2_ZN}
\eeq
Note that this expression is also divergent for $\eta \rightarrow 0$.

We will now describe two different prescriptions to define sum rules of order one, which are perfectly finite (without 
loss of generality we will use as reference domain the circle).

In the first approach one considers suitable linear combinations of Green's functions corresponding to different boundary conditions, 
in such a way that the divergent terms identically vanish. 

Looking at the example that we have just discussed, for instance, we 
see that the Green's functions for Dirichlet and for Neumann bc contain the same divergent term for $\eta \rightarrow 0$ and therefore
the combination 
\beq 
\Delta G^{(D/N)}(r,\theta,r + \eta \delta r,\theta+ \eta \delta\theta) &\equiv &
G^{(D)}(r,\theta,r + \eta \delta r,\theta+ \eta \delta\theta) - G^{(N)}(r,\theta,r + \eta \delta r,\theta+ \eta \delta\theta)   \nonumber \\
&\approx& \frac{3}{8 \pi }-\frac{r^2}{2 \pi } +\frac{\log \left(1-r^2\right)}{\pi } + O(\eta)
\label{2_GDN}
\eeq 
is finite for $\eta \rightarrow 0$ \footnote{Note that this combination does not depend on how $(r' ,\theta') \rightarrow (r,\theta)$.}.

As a result the corresponding trace
\beq
Z_{D/N}(1) &=& \sum_{n=1}^\infty \left( \frac{1}{E_n^{(D)}} -\frac{1}{E_n^{(N)}}\right) \nonumber \\
&=& \int_0^1 dr \int_0^{2\pi}d\theta \ r \ \Delta G^{(D/N)}(r,\theta,r + \eta \delta r,\theta+ \eta \delta\theta) 
  \Sigma(r,\theta) \nonumber \\
&+& \frac{\int_0^1 dr \int_0^{2\pi}d\theta \int_0^1 dr' \int_0^{2\pi}d\theta'  \ r r' \ G^{(N)}(r,\theta,r',\theta') \Sigma(r,\theta)\Sigma(r',\theta')}{\int_0^1 dr \int_0^{2\pi}d\theta \ r \ \Sigma(r,\theta)}
\eeq
is also finite. 

The second approach applies to domains with one (or more) symmetry axis: the eigenfunctions of the negative Laplacian on this domain will then be 
either be even or odd with respect to reflections about this axis and the eigenstates will be characterized also by a quantum number specifying to which 
symmetry class it belongs. In this case we can split the Green's function into an even an odd part with respect to reflexion about the symmetry axes;
in the case of the disk with Dirichlet boundary conditions at the border, for example, these Green's functions are
\beq
G^{(sym-D)}(r,\theta,r',\theta')  &=& \frac{1}{4} \left[ G^{(D)}(r,\theta,r',\theta') + G^{(D)}(r,-\theta,r',\theta') \right. \nonumber \\
&+& \left. G^{(D)}(r,\theta,r',-\theta') + G^{(D)}(r,-\theta,r',-\theta')\right] \nonumber \\
G^{(antisym-D)}(r,\theta,r',\theta')  &=& \frac{1}{4} \left[ G^{(D)}(r,\theta,r',\theta') - G^{(D)}(r,-\theta,r',\theta') \right. \nonumber \\
&-& \left. G^{(D)}(r,\theta,r',-\theta') + G^{(D)}(r,-\theta,r',-\theta')\right] \nonumber
\eeq

For $r'=r+ \eta \delta r$ and $\theta' = \theta +\eta \delta \theta$, with $\eta \rightarrow 0$, 
we introduce the linear combination
\beq
\Delta G_{S/A}^{(D)}(r,\theta)  &\equiv& \lim_{\eta \rightarrow 0} \left[ G^{(sym-D)}(r,\theta,r + \eta \delta r,\theta + \eta \delta\theta) 
- G^{(antisym-D)}(r,\theta,r + \eta \delta r,\theta + \eta \delta\theta) \right] \nonumber \\  
&\approx& \frac{\log \left(r^4-2 r^2 \cos (2 \theta )+1\right)-\log \left(4 r^2 \sin ^2(\theta)\right)}{4 \pi}+ O(\eta)
\eeq
that is perfectly finite for $\eta =0$.

As a result one has an explicit expression for the sum rule involving the eigenvalues of the even and odd states directly 
in terms of this Green's function as
\beq
Z_{D}^{(S/A)}(1) &=& \sum_{n=1}^\infty \left( \frac{1}{E_n^{(even-D)}} -\frac{1}{E_n^{(odd-D)}}\right)  \nonumber \\
&=& \int_0^1 dr \int_0^{2\pi} d\theta \ r \ \Delta G_{S/A}^{(D)}(r,\theta) \ \Sigma(r,\theta) 
\eeq

An analogous expression can be found for the case of Neumann bc.

\section{Applications}
\label{sec:appl}

In this section we consider several applications of the general formulas derived in section \ref{sec:sum_rules}, comparing the
exact results obtained with these formulas with the approximate results obtained numerically.

\subsection{Rectangle (Dirichlet bc)}

Consider a rectangle of sides $a$ and $b$ with Dirichlet bc at its borders. The expression for the Green's functions
corresponding to Dirichlet boundary conditions is reported in eqs.(8) and (9) of Ref.~\cite{Amore13B}.

Using that expression we work out the symmetrized and antisymmetrized Green's functions:
\beq
G^{(D-sym)}(x,y;x',y') &\equiv&  \frac{1}{4} \left[ G^{(D)}(x,y;x',y') + G^{(D)}(x,-y;x',y') \right. \nonumber \\
&+& \left. G^{(D)}(x,y;x',-y')+G^{(D)}(x,-y;x',-y')\right] \nonumber \\
G^{(D-antisym)}(x,y;x',y') &\equiv&  \frac{1}{4} \left[ G^{(D)}(x,y;x',y') - G^{(D)}(x,-y;x',y') \right. \nonumber \\
&-& \left. G^{(D)}(x,y;x',-y')+G^{(D)}(x,-y;x',-y')\right] \nonumber
\eeq
which correspond to selecting the contributions of even (odd) modes with respect to a symmetry axis.

We now define
\begin{eqnarray}
G^{(D)}_{S/A}(x,y;x',y') \equiv G^{(D-sym)}(x,y;x',y') -G^{(D-antisym)}(x,y;x',y') 
\end{eqnarray}
whose trace, by construction, is {\rm finite} and must be
\begin{eqnarray}
Tr \left[ G^{(D)}_{S/A} \right] = \sum_{n=1}^\infty \left[ \frac{1}{\epsilon^{(even)}_{n}} - \frac{1}{\epsilon^{(odd)}_{n}} \right]
\end{eqnarray}

Using the explicit expression for $G^{(D)}_{S/A}(x,y;x',y')$ one has
\beq
Tr \left[ G^{(D)}_{S/A} \right] &=&  \sum_{n=1}^\infty \int_{-a/2}^{a/2} dx  \int_{-b/2}^{b/2} dy  \frac{{\rm csch}\left(\frac{\pi  b n}{a}\right)
   \sin ^2\left(\frac{\pi  n (a+2 x)}{2 a}\right)}{\pi  n} \nonumber \\
&\cdot&   \left(\theta (-y) \left(\cosh \left(\frac{\pi  n (b+2 y)}{a}\right)-1\right)+\theta (y)
   \left(\cosh \left(\frac{\pi  n (b-2 y)}{a}\right)-1\right)\right) \nonumber \\
   &=& \sum_{n=1}^\infty \ \left[ \frac{a^2}{2 \pi ^2 n^2}-\frac{a b \ {\rm csch}\left(\frac{\pi  b n}{a}\right)}{2 \pi n} \right] \nonumber \\
   &=& \frac{a^2}{12} - \sum_{n=1}^\infty \frac{a b \ {\rm csch}\left(\frac{\pi  b n}{a}\right)}{2 \pi n}  \nonumber \\
   &=& \frac{a^2}{12} - \frac{a b}{\pi} \sum_{k=0}^\infty  \log \left(1-e^{-(2k+1)b\pi/a} \right)
\eeq
where the last series converges exponentially at large $k$.

For $a=b=\sqrt{2}$ the first ten terms of the series provide the sum rule exact to about $32$ digits:
\beq
Tr \left[ G^{(D)}_{S/A} \right] &\approx& 0.13849223337149623168244744136286\dots
\eeq

This result must be contrasted with the numerical result obtained by summing the explicit eigenvalues
\begin{eqnarray}
\sum_{n_x=1}^\infty \sum_{n_y=1}^\infty \left[ \frac{1}{((2 n_x-1)^2/a^2+n_y^2/b^2) \pi^2} - \frac{1}{((2 n_x)^2/a^2+n_y^2/b^2) \pi^2}\right]
\end{eqnarray}

When the sum is restricted to the lowest $5 \times 10^5$ eigenvalues assuming $a=b=\sqrt{2}$ one obtains
\beq
S_{500000} = \sum_{n=1}^{5\times 10^5} \left[ \frac{1}{\epsilon^{(even)}_{n}} - \frac{1}{\epsilon^{(odd)}_{n}} \right] \approx 
\underline{0.138}312\dots
\eeq
with just $3$ correct digits.

A simple fit, using the last $10^4$ partial sums, shows that the partial sums $S_N$ converge to the exact result as
\begin{eqnarray}
S_N \approx \underline{0.138492}0-\frac{0.1267811}{\sqrt{x}}
\end{eqnarray}
where the extrapolated value has now $6$ correct digits.

A still better estimate can be obtained approximating the tail of the series, $n> 5\times 10^5$, with the asymptotic behavior 
described by Weyl's law, which for $\lambda \gg 1$ states that the number of modes below the energy $\lambda$ is
\begin{eqnarray}
N(\lambda) &\approx& \frac{A \lambda}{4 \pi} -\frac{L \sqrt{\lambda}}{4 \pi} + \dots
\end{eqnarray}
where $A$ and $L$ are the area and perimeter of the domain respectively. In the present case $A=2$ and $L=4 \sqrt{2}$.

To approximate the tail of the series, one needs to obtain the leading asymptotic behavior of the odd and even modes; 
for the odd modes
\begin{eqnarray}
N^{(odd)}(\lambda) &\approx& \frac{A^{(odd)} \lambda}{4 \pi} -\frac{L^{(odd)} \sqrt{\lambda}}{4 \pi} + \dots
\end{eqnarray}
where 
\begin{eqnarray}
A^{(odd)} &=& \frac{1}{2} A = 1 \\
L^{(odd)} &=& \frac{3}{4} L = 3 \sqrt{2} 
\end{eqnarray}

Finally
\begin{eqnarray}
N^{(even)}(\lambda) &\approx& \frac{A^{(even)} \lambda}{4 \pi} -\frac{L^{(even)} \sqrt{\lambda}}{4 \pi} + \dots
\end{eqnarray}
where 
\begin{eqnarray}
A^{(even)} &=& A -A^{(odd)} = \frac{1}{2} A = 1 \\
L^{(even)} &=& L - L^{(odd)} = \frac{1}{4} L = \sqrt{2} 
\end{eqnarray}

Estimating the remainder of the series using the Weyl asymptotics we finally have
\beq
S_{500000} + \sum_{n=5\times 10^5+1}^\infty \left[ \frac{1}{\lambda_n^{(even)}} -\frac{1}{\lambda_n^{(odd)}} \right] \approx 
\underline{0.1384922}727
\eeq
where now $7$ digits have converged to their correct values.

\subsection{Disk (Dirichlet bc)}

The Helmholtz equation can be solved exactly when the domain is a unit disk, with Dirichlet boundary conditions
at the border; in this case the eigenfunctions
are simply given by
\beq
\Phi_{nms}(r,\theta) &=& N_{nm} J_m(\kappa_{nm} r) \times \left\{ \begin{array}{ccc}
\cos \theta & , & s=1\\
\sin \theta & , & s=2\\
\end{array}
\right.
\eeq
where $\kappa_{nm}$ is the $m^{th}$ zero of the Bessel function of order $n$ and $N_{nm}$ is a normalization 
constant.  The eigenvalues of the Helmholtz equation are also known and they are
\beq
E_{nms} = \left\{ \begin{array}{ccccc}
\kappa_{nm}^2 & , & s=1 & n=0,1,\dots & m=1,2,\dots \\
\kappa_{nm}^2 & , & s=2 & n=1,2,\dots & m=1,2,\dots \\
\end{array}
\right.
\eeq

From these formulas it is clear that the eigenfunctions corresponding to $s=1$ are even with respect to a
reflection about the horizontal axis, whereas the eigenfunctions corresponding to $s=2$ are odd.

In analogy to what we have done for the rectangle we define
\begin{eqnarray}
G^{(D-sym)}(r_1,\theta_1,r_2,\theta_2) &=& \frac{1}{4} \left[G^{(D)}(r_1,\theta_1,r_2,\theta_2)+
G^{(D)}(r_1,-\theta_1,r_2,\theta_2) \nonumber \right. \\ 
&+& \left. G^{(D)}(r_1,\theta_1,r_2,-\theta_2)+G^{(D)}(r_1,-\theta_1,r_2,-\theta_2)\right] \nonumber \\
G^{(D-antisym)}(r_1,\theta_1,r_2,\theta_2) &=& \frac{1}{4} \left[G^{(D)}(r_1,\theta_1,r_2,\theta_2)-
G^{(D)}(r_1,-\theta_1,r_2,\theta_2) \nonumber \right. \\ 
&-& \left. G^{(D)}(r_1,\theta_1,r_2,-\theta_2)+G^{(D)}(r_1,-\theta_1,r_2,-\theta_2)\right] \nonumber 
\end{eqnarray}
which correspond to selecting the contributions of the even and odd modes under reflections with respect to 
the $x$ axis.

As we have seen earlier, the Green's function obtained as
\beq
\Delta G^{(D)}_{S/A}(r_1,\theta_1,r_2,\theta_2) &\equiv& G^{(D-sym)}(r_1,\theta_1,r_2,\theta_2)
-G^{(D-antisym)}(r_1,\theta_1,r_2,\theta_2) \nonumber \\
&=& \frac{1}{4 \pi} \ \log \left(\frac{-2 r_1 r_2 \cos \left(\theta _1+\theta _2\right)+r_1^2
   r_2^2+1}{-2 r_2 r_1 \cos \left(\theta _1+\theta _2\right)+r_1^2+r_2^2}\right)
\eeq 
is free of divergencies when the limit $\theta_2 \rightarrow \theta_1$ and $r_2 \rightarrow r_1$ is taken
\beq
\lim_{r_2 \rightarrow r_1} \lim_{\theta2 \rightarrow \theta_1} \Delta  G^{(D)}_{S/A}(r_1,\theta_1,r_2,\theta_2)  = 
\frac{1}{4 \pi } \left[\log \left(r_1^4-2 r_1^2 \cos (2 \theta_1 )+1\right)-\log \left(4 r_1^2 \sin^2(\theta_1)\right) \right]
\eeq

The sum rule of order one 
\begin{eqnarray}
\sum_n \left[\frac{1}{\epsilon_n^{(even)}} - \frac{1}{\epsilon_n^{(odd)}}\right] =
\sum_{m=1}^\infty \frac{1}{\kappa_{0m}^2}
\end{eqnarray}
can thus be obtained by means of the trace
\beq
Tr \left[G^{(D)}_{S/A}\right] &=& \int_0^1 dr \ \int_0^{2\pi} d\theta \ r \ \Delta G^{(D)}_{S/A}(r,\theta,r,\theta) \nonumber \\
&=& \int_0^1 dr \ \int_0^{2\pi} d\theta \  \frac{r}{4 \pi } 
    \left( \log \left(r^4-2 r^2 \cos (2 \theta )+1\right)-\log \left(4 r^2 \sin ^2(\theta )\right)\right) \nonumber \\
&=& \frac{1}{4} \label{sum_circleD}
\eeq

This sum rule is a particular case of the "circular partial wave zeta function" calculated by Steiner~\cite{Steiner87} (see also Ref.~\cite{Elizalde93}):
\beq
\zeta_l(s) \equiv \sum_{m=1}^\infty \frac{1}{\kappa_{lm}^{2s}}
\eeq
that, for $s=1$, (see Eq.(5.18) of ref.~\cite{Steiner87}) yields 
\beq
\zeta_l(1) = \frac{1}{4 (l+1)}
\label{zeta_steiner}
\eeq
The special case $l=0$ corresponds to the sum rule considered here and it agrees with the result of Eq.~(\ref{sum_circleD}).

\subsection{Disk (Neumann bc)}

For the case of Neumann bc we introduce the Green's functions with appropriate parity
\begin{eqnarray}
G^{(N-sym)}(r_1,\theta_1,r_2,\theta_2) &=& \frac{1}{4} \left[G^{(N)}(r_1,\theta_1,r_2,\theta_2)+
G^{(N)}(r_1,-\theta_1,r_2,\theta_2) \nonumber \right. \\ 
&+& \left. G^{(N)}(r_1,\theta_1,r_2,-\theta_2)+G^{(N)}(r_1,-\theta_1,r_2,-\theta_2)\right] \nonumber \\
G^{(N-antisym)}(r_1,\theta_1,r_2,\theta_2) &=& \frac{1}{4} \left[G^{(N)}(r_1,\theta_1,r_2,\theta_2)-
G^{(N)}(r_1,-\theta_1,r_2,\theta_2) \nonumber \right. \\ 
&-& \left. G^{(N)}(r_1,\theta_1,r_2,-\theta_2)+G^{(N)}(r_1,-\theta_1,r_2,-\theta_2)\right] \nonumber 
\end{eqnarray}
and use them to define
\beq
\Delta G^{(N)}_{S/A}(r_1,\theta_1,r_2,\theta_2) &\equiv& G^{(N-sym)}(r_1,\theta_1,r_2,\theta_2)
-G^{(N-antisym)}(r_1,\theta_1,r_2,\theta_2) \nonumber \\
&=& -\frac{\log \left(\left(-2 r_2 r_1 \cos \left(\theta _1+\theta
   _2\right)+r_1^2+r_2^2\right) \left(r_1 r_2 \left(r_1 r_2-2 \cos \left(\theta
   _1+\theta _2\right)\right)+1\right)\right)}{4 \pi }\nonumber \\ 
&+& \frac{r_1^2}{4 \pi}+\frac{r_2^2}{4 \pi }-\frac{3}{8 \pi }
\eeq

In the limit $\theta_2 \rightarrow \theta_1$ and $r_2 \rightarrow r_1$ this expression reduces to
\beq
\lim_{r_2 \rightarrow r_1} \lim_{\theta_2 \rightarrow \theta_1} \Delta G^{(N)}_{S/A}(r_1,\theta_1,r_2,\theta_2)  = 
-\frac{\log \left(4 r_1^2 \sin ^2\left(\theta _1\right) \left(-2 r_1^2 \cos \left(2
   \theta _1\right)+r_1^4+1\right)\right)}{4 \pi }+\frac{r_1^2}{2 \pi }-\frac{3}{8 \pi }
\eeq
and once again it is free of divergencies.

The sum rule
\begin{eqnarray}
\sum_n \left[\frac{1}{\epsilon_n^{(N-even)}} - \frac{1}{\epsilon_n^{(N-odd)}}\right] \rightarrow
\sum_{m=1}^\infty \frac{1}{\rho_{0m}^2}
\end{eqnarray}
where $\rho_{nm}$ is the $m^{th}$ zero of the derivative of the Bessel function of order $n$, can 
then be obtained calculating the trace
\beq
Tr \left[\Delta G^{(N)}_{S/A}\right] &=& \int_0^1 dr \ \int_0^{2\pi} d\theta \ r \  \Delta G^{(N)}_{S/A}(r,\theta,r,\theta) 
= \frac{1}{8}
\eeq

Using the property $\frac{d}{dx} J_0(x) = - J_1(x)$ we have that $\rho_{0,m} = \kappa_{1,m}$ and therefore 
the sum rule is a special case of  eq.~(\ref{zeta_steiner}), for $l=1$:
\beq
\zeta_1(1) = \frac{1}{8}
\eeq
confirming our result.

\subsection{Disk  (Dirichlet-Neumann bc) }

Still another possibility is to consider the sum rule 
\beq
\sum_{n=1}^\infty \left[ \frac{1}{\epsilon_n^{(D)}} - \frac{1}{\epsilon_n^{(N)}}\right]
\label{SR_disk_DN}
\eeq
where the Neumann zero mode ($n=0$) is excluded from the sum.

In this case
\beq
\Delta G^{(D/N)}(r,\theta,r,\theta)  = G^{(D)}(r,\theta,r,\theta) -G^{(N)}(r,\theta,r,\theta)  = \frac{3}{8\pi} - \frac{r^2}{2\pi} + \frac{1}{\pi}\log(1-r^2)
\eeq
and its trace is
\beq
Tr \left[\Delta G^{(D/N)}\right] &=& \int_0^1 dr \ \int_0^{2\pi} d\theta \ r \left[ \frac{3}{8\pi} - \frac{r^2}{2\pi} + \frac{1}{\pi}\log(1-r^2)\right] \nonumber \\
&=& -\frac{7}{8}
\eeq

It is useful to introduce the appropriate set of "circular partial wave zeta function" for Neumann bc
\beq
\eta_l(s) \equiv \sum_{m=1}^\infty \frac{1}{\rho_{lm}^{2s}}
\eeq
extending the definitions of Steiner \cite{Steiner87}.

Based on accurate numerical calculations we have been able to guess the general expression for $\eta_l(1)$, 
which reads
\beq
\eta_l(1) &=& \left\{ \begin{array}{ccc}
\frac{1}{8} & , & l=0 \\
\frac{l+2}{4 l (l+1)} & , & l \geq 1 \\
\end{array} \right.
\eeq
where $\eta_0(1) = \zeta_1(1)$ has been used in the previous case.

By substituting these expressions into the series (\ref{SR_disk_DN}) we have
\beq
\sum_n \left[ \frac{1}{\epsilon_n^{(D)}} - \frac{1}{\epsilon_n^{(N)}}\right] &=& 
\left(\frac{1}{4}-\frac{1}{8}\right) + 2\sum_{l=1}^\infty \left( \frac{1}{4 (l+1)}- \frac{l+2}{4 l (l+1)} \right)   \nonumber \\
&=& -\frac{7}{8}
\eeq
which confirms the result obtained with the trace.

\subsection{Circular annulus (Dirichlet bc)}

Following ref.~\cite{Amore11} we consider a rectangle of sides $a= -\log r_0$ and $b=2 \pi$. The conformal map
\beq
g(z) = e^{z+1/2\log r_0}
\label{map_annulus}
\eeq
transforms the rectangle $(|x| \leq a/2,|y| \leq b/2)$ into an annulus of radii $r_{min} = r_0$ and $r_{max} = 1$ ($r_0<1$). 

The "conformal density" associated with this transformation is 
\beq
\Sigma(x,y) = r_0 e^{2x}
\eeq

By taking Dirichlet and periodic boundary conditions along the horizontal and vertical sides of the rectangle respectively,  
the annulus resulting from the conformal map obeys Dirichlet boundary conditions at the border. 
As a result, we  need to use the Green's function for a rectangle with Dirichlet-periodic boundary conditions reported in 
ref.~\cite{Amore13B}.

The sum rule
\beq
\sum_{n=1}^{\infty} \left[ \frac{1}{\epsilon_n^{(even)}}-\frac{1}{\epsilon_n^{(odd)}}\right]
\eeq
can then be expressed as
\beq
\mathcal{S} &=& Tr \left\{ \left[ G^{(DP-even)} -G^{(DP-odd)} \right] \Sigma \right\} \nonumber \\
&=& \sum_{n_x=1}^\infty    \int_{-a/2}^{a/2} dx \ \int_{-b/2}^{b/2} dy
\left[ g_{n_y,1}^{(DP)}(x,x) \left( \phi^{(P)}_{n_y,1}(y)\right)^2 - g_{n_y,2}^{(DP)}(x,x) \left( \phi^{(P)}_{n_y,2}(y)\right)^2\right]
\Sigma(x,y) \nonumber \\
&+&  \int_{-a/2}^{a/2} dx \ \int_{-b/2}^{b/2} dy \ g_{0,1}^{(DP)}(x,x) \left(\phi^{(P)}_{0,1}(y)\right)^2
\eeq

One can carry out the calculation explicitly, and see that the first contribution identically vanishes; the final result is
particularly simple
\beq
\mathcal{S} = \frac{1}{4} \left(1+r_0^2+\frac{1-r_0^2}{\log \left(r_0\right)}\right)
\eeq

It is interesting to discuss the limits $r_0 \rightarrow 0^+$ and $r_0 \rightarrow  1^-$:
\beq
\lim_{r_0\rightarrow 0^+} \mathcal{S} &=& \frac{1}{4}+\frac{1}{4 \log\left(r_0\right)}+\left(\frac{1}{4}-\frac{1}{4 \log\left(r_0\right)}\right) r_0^2 + \dots \\
\lim_{r_0\rightarrow 1^-} \mathcal{S} &=& \frac{1}{6} \left(-1+r_0\right)^2 + \dots
\eeq

In the first case the sum rule tends to the corresponding sum rule for the disk (although the expression is non-analytical at $r=r_0$); in the second
case the sum rule tends to zero, because of the transversal modes whose energy grows without bounds as the transversal size of the annulus is shrinked to zero.

This sum rule can be tested using the explicit form for the solution of the Helmholtz equation:
\beq
\Phi_{n,m,s}(r,\theta) &=& N_{nms} \left[ Y_n(k_{nm}) J_n(k_{nm} r/r_0)-J_n(k_{nm}) Y_n(k_{nm} r/r_0) \right] \nonumber \\
&\times& \left\{ 
\begin{array}{ccc}
\cos n\theta &,& s=1 \\
\sin n\theta &,& s=2 \\
\end{array}
\right.
\eeq
where $k_{nm}$ are the roots of the equation
\beq
 Y_n(k_{nm}) J_n(k_{nm}/r_0)-J_n(k_{nm}) Y_n(k_{nm}/r_0) = 0
\label{eq_zero}
\eeq
and the corresponding eigenvalues are
\beq
E_{nm} = \frac{k_{nm}^2}{r_0^2}
\eeq

In analogy with the case of the disk (see ref.~\cite{Steiner87}) one can define an "annular partial wave zeta function" as
\beq
\tilde{\zeta}_l(s) \equiv \sum_{m=1}^\infty \frac{r_0^2}{k_{lm}^{2s}}
\eeq
and express the sum rule as
\beq
\mathcal{S} = \tilde{\zeta}_0(1) 
\eeq

We have calculated numerically the first $500$ roots of eq.~(\ref{eq_zero}) for $r_0 = 1/2^j$ and $j=1,\dots, 10$.
The results are reported in Table \ref{table1} and show that almost all the digits of the exact results are reproduced
when the tail of the series is estimated using the Weyl asymptotics.

\begin{table}[tbp]
\caption{Dirichlet sum rules for a circular annulus at different values of $r_0$. The column labeled
"numerical" corresponds to the approximate sum rules obtained using the lowest $500$ roots of eq.~(\ref{eq_zero}).}
\bigskip
\label{table1}
\begin{center}
\begin{tabular}{|c|c|c|c|}
\hline
$r_0$ & $exact$ & $numerical$  & $numerical + Weyl$ \\
\hline
$1/2$    & $0.04199467983$ & $0.04194406987$ & $0.04199467985$ \\
$1/4$    & $0.09655917490$ & $0.09644530247$ & $0.09655917493$ \\
$1/8$    & $0.1355601724$  & $0.1354051794$  & $0.1355601725$  \\
$1/16$   & $0.1611603429$  & $0.1609824172$  & $0.1611603430$  \\
$1/32$   & $0.1781798327$  & $0.1779898474$  & $0.1781798327$  \\
$1/64$   & $0.1899634176$  & $0.1897672542$  & $0.1899634177$  \\
$1/128$  & $0.1984935807$  & $0.1982942908$  & $0.1984935807$  \\
$1/256$  & $0.2049202826$  & $0.2047194195$  & $0.2049202824$  \\
$1/512$  & $0.2099262443$  & $0.2097245911$  & $0.2099262433$  \\
$1/1024$ & $0.2139328968$  & $0.2137308457$  & $0.2139328943$  \\
\hline
\end{tabular}
\end{center}
\bigskip\bigskip
\end{table}

\subsection{Cardioid-like region (Dirichlet bc)}

Following \cite{Robnik83,Robnik84} we consider the conformal map
\beq
g(z) = \frac{z+ \lambda z^2}{\sqrt{1+2\lambda^2}} \ \ \ ; \ \ \  0 \leq \lambda \leq 1/2
\eeq
which maps the unit disk to a cardioid-like region, leaving the area constant (the particular choice $\lambda=1/2$ 
corresponds to a cardioid).

The original interest of refs.~\cite{Robnik83,Robnik84} into studying these
domains is related to the fact that the cardioid is known to exhibit
quantum chaos and its level spacing distribution is in good agreement with a 
Wigner distribution.

The spectrum and eigenfunctions for this problem, unlike in the cases that we have studied earlier, 
are not known, and intensive numerical calculations are required to obtain good approximations
to the lower part of the spectrum~\cite{Backer95}. Being able to obtain an exact sum rule in this problem can then be 
valuable to assess the quality of the numerical calculations and possibly extract information on the subleading terms 
of the asymptotic behavior of the spectrum as well.

The conformal density in this case is
\begin{eqnarray}
\Sigma(r,\theta) = \frac{4 \lambda ^2 r^2+4 \lambda  r \cos (\theta )+1}{2 \lambda ^2+1}
\end{eqnarray}
and we may calculate the sum rule
\beq
\sum_{n=1}^\infty \left[ \frac{1}{\epsilon_n^{(D-even)}}-\frac{1}{\epsilon_n^{(D-odd)}}\right]
\eeq
using  the trace
\beq
Tr \left[G^{(D)}_{S/A} \Sigma \right] &=& \int_0^1 dr \ \int_0^{2\pi} d\phi \ r  \ \Sigma(r,\theta)
G^{(D)}_{S/A}(r,\theta,r,\theta) \nonumber \\
&=& \frac{1+ \lambda^2}{4 (1+2 \lambda ^2)}
\eeq

This result can be compared with the approximate result obtained calculating numerically a large number of
even and odd eigenvalues, using the Rayleigh-Ritz method. For the even (odd) states we have worked with a set
of $11984$ ($11886$) states, which are sufficient to obtain the lowest $5000$ eigenvalues in each symmetry
class precisely. In Table \ref{table2} we compare the exact sum rules (second column) with the partial sums 
over the first 5000 numerical eigenvalues (third column) and with the partial sums where the contribution
of the tail of the series is estimated using Weyl's asymptotics (fourth column).
These last results agree with the exact one to 5 digits.

\begin{table}[tbp]
\caption{Dirichlet sum rules for cardioid--like regions at different values of $\lambda$. The column labeled
"numerical" corresponds to the approximate sum rules obtained using the lowest $5000$ eigenvalues.}
\bigskip
\label{table2}
\begin{center}
\begin{tabular}{|c|c|c|c|}
\hline
$\lambda$ & $exact$ & $numerical$  & $numerical + Weyl$ \\
\hline
$0.0$ & $0.25000000$ & $0.24683904$ & $0.249999954$\\
$0.1$ & $0.24754902$ & $0.24441922$ & $0.247548921$\\
$0.2$ & $0.24074074$ & $0.23769926$ & $0.240740593$\\
$0.3$ & $0.23093220$ & $0.22802280$ & $0.230932079$\\
$0.4$ & $0.21969697$ & $0.21694770$ & $0.219697894$\\
$0.5$ & $0.20833333$ & $0.20575859$ & $0.208337812$\\
\hline
\end{tabular}
\end{center}
\bigskip\bigskip
\end{table}

\subsection{Cardioid-like region (Dirichlet - Neumann bc)}

In this case we consider the sum rule relative to the difference between Dirichlet and Neumann eigenvalues:
\beq
\sum_{n=1}^\infty \left[ \frac{1}{\epsilon_n^{(D)}}-\frac{1}{\epsilon_n^{(N)}}\right]
\eeq

The expression of this sum rule in terms of the Green's functions is
\beq
Tr\left[( G^{(D)}-G^{(N)}) \Sigma \right] &+& \frac{\int r_1 r_2 \Sigma(r_1,\theta_1) G^{(N)}(r_1,\theta_1,r_2,\theta_2) \Sigma(r_2,\theta_2) 
dr_1 dr_2 d\theta_1 d\theta_2}{\int \Sigma r dr d\theta} \nonumber \\
&\equiv& S_1 + S_2
\eeq
where the second term corresponds to eq.(19) of ref.~\cite{Amore14} \footnote{Note that this term is absent in the disk because the "regularized"
Neumann Green's function is orthogonal to the zero mode.}.

The first term can be calculated explicitly
\beq
S_1 &=& \int_0^1 dr \int_0^{2\pi} d\theta \frac{r}{8 \pi  \left(2 \lambda ^2+1\right)}  \left(-4 r^2+\log \left(\left(r^2-1\right)^8\right)+3\right) \left(4 \lambda ^2
   r^2+4 \lambda  r \cos (\theta )+1\right)\nonumber \\
   &=& -\frac{7 \left(3+10 \lambda ^2\right)}{24 (1+ 2 \lambda^2)}
\eeq

The expression for $S_2$ is much more involved and it reads
\beq
S_2 &=&  \int_0^1 dr_1 \int_0^{2\pi} d\theta_1 \int_0^1 dr_2 \int_0^{2\pi} d\theta_2 
\frac{r_1 r_2}{2 \pi ^2 \left(2 \lambda ^2+1\right)^2}  \nonumber \\
&\times&
\left(4 \lambda  r_1 \cos \left(\theta _1\right)+4 \lambda ^2 r_1^2+1\right) 
\left(4 \lambda  r_2 \cos \left(\theta _2\right)+4 \lambda ^2 r_2^2+1\right)\nonumber \\
   &\times&
 \left[-\log \left(\sqrt{-2 r_2 r_1 \cos \left(\theta _1-\theta_2\right)+r_1^2+r_2^2}\right)
       -\log \left(\sqrt{-2 r_1 r_2 \cos \left(\theta _1-\theta_2\right)+r_1^2 r_2^2+1}\right) \right. \nonumber \\
  &+& \left. \frac{1}{2} \left(r_1^2+r_2^2\right)-\frac{3}{4}\right]
\eeq

For the case of the cardioid the sum rule provides
\beq
S_1+S_2 = - \frac{77}{72}+0.1320297676 = -0.9374146768
\eeq

We have also calculated numerically the eigenvalues of the cardioid using the collocation method described in ref.~\cite{Amore10}
both for Dirichlet and Neumann bc. Using grids of about $80000$ points we have obtained the first $10^4$ eigenvalues with good accuracy.
The partial sum obtained in this case is
\beq
\sum_{n=1}^{10000} \left[ \frac{1}{\epsilon_n^{(D)}}-\frac{1}{\epsilon_n^{(N)}}\right] = -0.927038
\eeq

This result can be improved estimating the tail of the series using the Weyl asympotic for Dirichlet and Neumann modes:
\beq
\sum_{n=1}^{10000} \left[ \frac{1}{\epsilon_n^{(D)}}-\frac{1}{\epsilon_n^{(N)}}\right] +
\sum_{n=10001}^{\infty} \left[ \frac{1}{\epsilon_n^{(Weyl-D)}}-\frac{1}{\epsilon_n^{(Weyl-N)}}\right] = -0.937038
\eeq

In this case the inferior accuracy of the collocation method, compared with the Rayleigh-Ritz method used earlier, reflects in the 
precision of the numerical sum rule, which agrees only to three digits with the exact result.


\section{Conclusions}
\label{sec:concl}

The eigenvalues of the negative laplacian on a finite two dimensional domain, subject to the appropriate boundary conditions,
can be used to define the spectral sum rules of eq.~(\ref{spectral_sum}). In refs.~\cite{Amore13A,Amore13B, Amore14} we have
derived general expressions for the sum rules of order $p$, with $p=2,3,\dots$; in the present paper we have considered the 
special case corresponding to $p=1$, for which the expressions of refs.~\cite{Amore13A,Amore13B, Amore14} diverge in two or more dimensions. 
This behavior is understood taking into account the asymptotic growth of the eigenvalues, given
by Weyl's law. We show that it is possible to cure this divergence, obtaining sum rules that are  finite and well defined, by taking 
suitable linear combinations of the divergent sum rules. The linear combinations should be such that the divergent term in the Green's function
appearing in the trace identically vanish. This can be accomplished in different ways, for instance building a sum rule out of Dirichlet and Neumann
eigenvalues, or out of eigenvalues belonging to different symmetry classes. We have worked out explicit expressions for a number of 
cases and compared them with the approximate results obtained with a direct numerical evaluation of a large number of eigenvalues (in this respect these 
sum rules provide a tool to assess the efficiency of the numerical methods used).  

An accurate calculation of the sum rules starting from the eigenvalues implies, apart from a reliable and precise numerical estimate of the 
lowest eigenvalues (say few thousands of them), also a good knowledge of the behavior of the higher part of the spectrum. As the quality of
the numerical estimates of the lower part of the spectrum improves, one may hope to obtain interesting information on the subleading terms in
the behavior of the higher part of the spectrum, by comparing the numerical results with the exact ones. It is important to stress that
the spectral sum rules of order one offer at least two advantages over the sum rules of higher orders: it is much easier to calculate them 
explicitly and they are more sensitive to the asymptotic behavior of the spectrum. In our opinion these features make them an intersting tool
for investigating the high frequency behavior of the spectrum.


\section*{Acknowledgements}
The research of P.A. was supported by the Sistema Nacional de Investigadores (M\'exico).



\end{document}